\documentclass[sigconf,natbib=false,anonymous=false]{acmart}

\AtBeginDocument{%
  }

\setcopyright{acmlicensed}
\copyrightyear{2025}
\acmYear{2025}
\acmDOI{https://doi.org/10.1145/3746252.3760932}

\acmConference[CIKM '25]{The 34th ACM International Conference on Information and Knowledge Management}{November 10--14,
  2025}{Seoul, Korea}

\acmISBN{978-1-4503-XXXX-X/2018/06}

\RequirePackage[
  datamodel=acmdatamodel,
  style=acmnumeric,
  ]{biblatex}

\begin{document}

\title{Ultra Fast Warm Start Solution for Graph Recommendations}

\author{Viacheslav Yusupov}
\affiliation{%
  \institution{HSE University}
  \city{Moscow}
  \country{Russian Federation}
}
\email{v.yusupov.lab@gmail.com}

\author{Maxim Rakhuba}
\affiliation{%
  \institution{HSE University}
  \city{Moscow}
  \country{Russian Federation}}

\author{Evgeny Frolov}
\affiliation{%
  \institution{AIRI}
  \city{Moscow}
  \country{Russian Federation}
}
\affiliation{%
  \institution{HSE University}
  \city{Moscow}
  \country{Russian Federation}
}

\renewcommand{\shortauthors}{Yusupov et al.}

\begin{teaserfigure}
    \centering
    \includegraphics[width=0.77\linewidth]{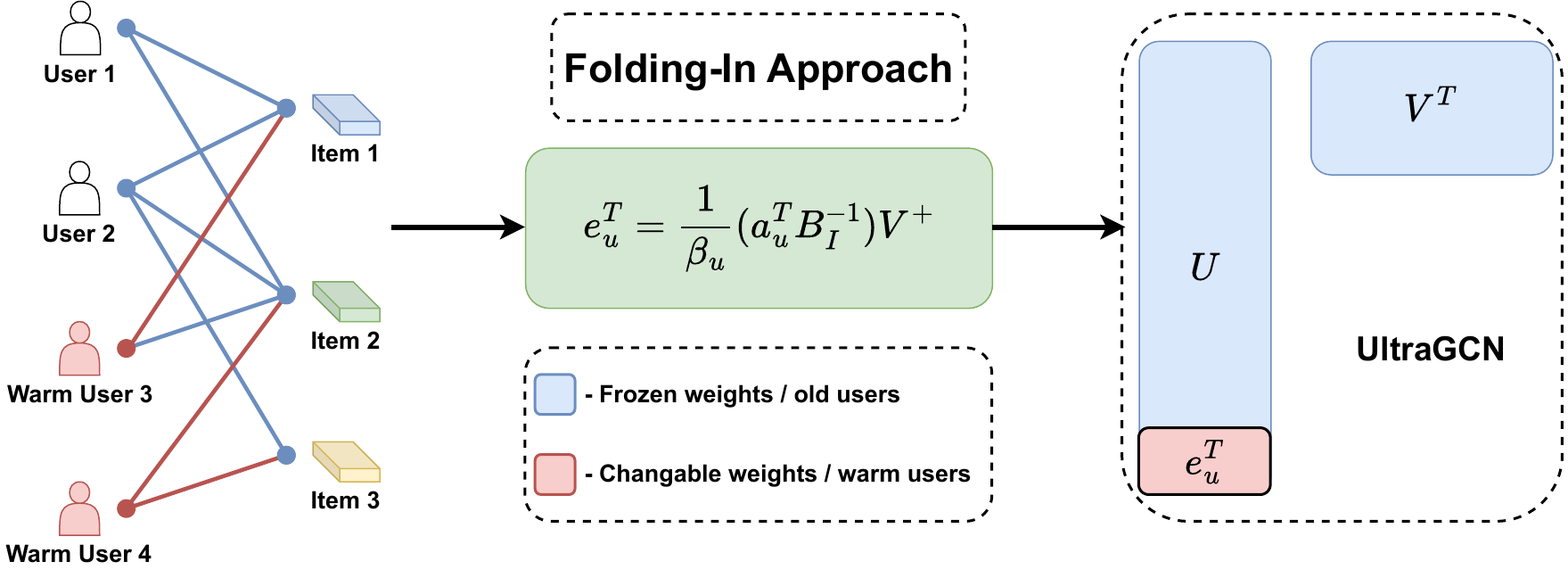}
    \caption{The  image illustrates the warm-start scenarios, with warm users—those who have new interactions—marked in red, and others in blue. The scheme on the right presents the Folding-In approach for the UltraGCN \cite{mao2021ultragcn} graph-based recommender system. The user embedding matrix is denoted as $U$, and the item embedding matrix as $V$. In the solution, $a_u$ is the vector of user $u$ interactions, the element $\beta_u$ and the diagonal matrix $B_I$ contain information about the graph structure and $V^+$ denotes a pseudo-inversed matrix. Trainable user embeddings ($e_u$) are shown in red, while frozen embeddings are shown in blue.}
    \label{fig:main}
\end{teaserfigure}

\vspace{-3pt}
\begin{abstract}
 In this work, we present a fast and effective Linear approach for updating recommendations in a scalable graph-based recommender system UltraGCN. Solving this task is extremely important to maintain the relevance of the recommendations under the conditions of a large amount of new data and changing user preferences. To address this issue, we adapt the simple yet effective low-rank approximation approach to the graph-based model. Our method delivers instantaneous recommendations that are up to $30$ times faster than conventional methods, with gains in recommendation quality, and demonstrates high scalability even on the large catalogue datasets.
\end{abstract}

\vspace{-3pt}


\vspace{-2pt}

\maketitle

\vspace{-5pt}
\section{Introduction}
Recommender systems play a crucial role in modern digital life; however, they face significant challenges due to a high information overload. To maintain the relevance of recommendations, it is essential that recommendation models are regularly updated. As the complexity of these models and the number of users continue to grow, the traditional model retraining becomes an inadequate solution for instant recommendations updates.

To address this issue, it is important to focus on updating recommendations only for those users who have interacted since the model was last trained or to tackle the \emph{warm start} problem \cite{warm} using the \emph{Folding-In} approach. Users with new interactions are referred to as \emph{warm users} (see Figure \ref{fig:main}). In recommender systems, sequential-based models \cite{sasrec,sasrecce,bert4rec} and graph-based models \cite{wang2019neural,he2020lightgcn,mao2021ultragcn} have shown outstanding performance in various domains.

Recent advances in sequential-based models \cite{attention} have established a new state-of-the-art result in the field of sequential recommendations \cite{sasrecce}. These models are fast and effective, effectively addressing the warm start problem by naturally updating user embeddings based on their interaction history \cite{sasrec}. However, sequential-based models may not yield significant benefits for certain datasets, particularly in cases where sequential patterns are absent \cite{klenitskiy2024does}.

In contrast, graph-based recommender systems \cite{he2020lightgcn, mao2021ultragcn} learn the global information about user-item interactions according to the interaction graph. These models demonstrate high recommendation quality, capturing the complex dependencies in the data. However, graph approaches suffer from high computational demands and the necessity of updating user representations to maintain the reliability of recommendations \cite{gao2023survey,zhang2024linear}. For this reason, it is crucial for graph-based models to develop Folding-In approaches to update user embeddings in the warm start scenario \cite{jorge2010latent}.

The existing Stochastic Gradient Descent-based Folding-In approaches \cite{sgd} are universal but slow and computationally consuming. To overcome this problem, we developed a specialized Folding-In approach \emph{Linear} for the simple and commonly used UltraGCN recommender system \cite{mao2021ultragcn}. Compared to previous fine-tuning \cite{yang2024graphpro} and meta-learning \cite{zhang2020retrain} approaches, our method updates only one warm user embedding and could be applied on the fly with up to $30
\times$ speed-up and better quality of recommendations compared to conventional SGD or fine-tuning approaches. In addition, to demonstrate the effectiveness and linear scalability \cite{singh2020scalability,roy2022systematic} of our Linear method of updating recommendations, we tested it on datasets with large catalogues. Furthermore, our approach shows a higher diversity of recommendations due to the exact solution for updated embeddings, as well as the small impact on popularity bias \cite{chen2023bias}.

Overall, our contributions are as follows.
\begin{itemize}
     \item We developed a novel fast and efficient Folding-In approach Linear for graph-based recommender systems, achieving better quality than conventional SGD-based approaches with significantly $30\times$ faster inference with better quality of recommendations.
     \item We theoretically and empirically study the scalability law of our Folding-In method.
     \item We demonstrate the effectiveness of our approach to increase the coverage of item catalogue.
 \end{itemize}
The rest of the paper is organized as follows: Section \ref{related} contains the related work,  Section \ref{prel} provides an explanation of UltraGCN and SVD models, Section \ref{method} details our approach, Sections \ref{exp}, \ref{res}, and \ref{conc} contain our experimental setup, results, and conclusion, respectively. The code for reproducing our results is available\footnote{https://github.com/YusupovV-Lab/UltraFastFoldIn}.

\vspace{-3pt}
\section{Related Work}
\label{related}
Graph-based recommender systems \cite{wang2019neural,mao2021ultragcn,he2020lightgcn,gao2023survey} provide high-quality recommendations by leveraging the graph structure of the data and employing graph neural networks \cite{gnn,kipf2016semi}. Despite their effectiveness, training these systems often requires significant amounts of time and computational resources \cite{zhang2024linear}. Consequently, retraining the model to capture new data becomes impractical in an online context.
Some methods, such as conventional SGD-based approaches \cite{sgd}, are general but do not fully utilize the graph structure inherent in the model. In contrast, techniques such as metalearning \cite{zhang2020retrain} and graph prompting \cite{yang2024graphpro,yi2023contrastive,sun2022gppt} update the entire model, which can be resource-intensive and time consuming.
To address the limitations of these approaches, we have developed a fast and effective Folding-In approach for the UltraGCN model \cite{mao2021ultragcn}, which allows real-time updates of recommendations during inference and updates only several embeddings of users with a changed history of interactions.

\vspace{-3pt}
\section{Preliminaries}
\label{prel}
In this section, we provide some preliminaries of our approach and discuss the methods utilized in this work. In the following sections $N$ denotes the number of items in the catalogue, $M$ the number of users, and $d$ is the size of the embeddings.

\subsection{Folding-In for SVD}
\label{svd}
First, we start with a traditional matrix factorization approach \cite{puresvd}, which solves the following problem:
\begin{equation}
    \min_{B:\text{ }rank(B) \le d}\|A_0 - B\|_F^2 = \|A_0 - U\Sigma V^T\|_F^2
    \label{task}
\end{equation}
where $A_0 \in \mathbb{R}^{M \times N}$ is the binary user-item interaction matrix and $U \in \mathbb{R}^{M\times d}$, $\Sigma \in \mathbb{R}^{d \times d}$ and $V\in \mathbb{R}^{N \times d}$ are the components of the SVD decomposition of $A_0$. The matrix $U$ has the meaning of user embeddings and $V\Sigma$ is an item embedding matrix.

When $A_0$ changes rapidly through new interactions, model retraining is infeasible due to its computational cost. To overcome this problem, we can update only the embeddings of warm users who have new interactions\cite{ekstrand2011collaborative, puresvd}. In this case, the optimization problem is formulated as follows:
\begin{equation}
\|a_u^T - e_u^T\Sigma V^T\|_2^2 \rightarrow \operatorname*{min}_{e_u},
\label{fold}
\end{equation}
where $e_u$ is the embedding of the user $u$  and the corresponding row of the matrix $U$. Despite the fact that this problem could be minimized by iterative approaches, such as SGD \cite{sgd}, the exact solution to the problem \eqref{fold} be obtained explicitly using the orthogonality of $V$. The final solution for user $u$ embedding has the $O(Nd)$ time and memory complexity:
\begin{equation}
    e_u^T = (a_u^TV)\Sigma^{-1}.
\end{equation}
Therefore, this folding-in approach provides fast and exact solutions without approximation errors and convergence issues as in optimization methods \cite{sgd,bfgs}.

\vspace{-3pt}
\subsection{Ultra Simplified Graph Neural Network}
\label{ultra}
To test our Folding-IN approach, we utilize the efficient UltraGCN \cite{mao2021ultragcn} model, which is the simplification of GCN-based recommender systems \cite{he2020lightgcn,wang2019neural}. In this model, the user-item interaction matrix is simplified to two vectors $b_U \in \mathbb{R}^M$ and $b_I \in \mathbb{R}^N$, collecting information about the total number of interactions for each user and item, respectively. 
Vectors $b_U$ and $b_I$ consist of elements $\beta_u$ and $\beta_i$, where:
\begin{equation}
    \beta_u = \frac{\sqrt{d_u+1}}{d_u} \text{ and } \beta_i = \frac{1}{\sqrt{d_i+1}}.
    \label{betas}
\end{equation}
Additionally, we introduce the matrices $B_U = diag(b_U)$ and $B_I = diag(b_I)$ for shorter notation. The values $d_u$ and $d_i$ are the degrees of vertices in the interaction graph, corresponding to the user $u$ and the item $i$, respectively. The score function of the model is:
\begin{equation}
    r_{ui} = \beta_u\beta_i e_u^Te_i, 
    \label{score}
\end{equation}
where $e_u \in \mathbb{R}^d$ and $e_i\in \mathbb{R}^d$ are user and item embeddings. The model is trained with a loss function combined from the different BPR components \cite{bpr} $\mathcal{L} = \mathcal{L}_B + \lambda \mathcal{L}_O$. In this expression:
\begin{equation}
    \begin{aligned}
       & \mathcal{L}_B =  -\sum_{u = 1}^M\sum_{i \in \mathcal{N}_u}\sum_{j \notin \mathcal{N}_u} ln(\sigma(\beta_{u}\beta_{i}e_u^Te_i - \beta_{u}\beta_{j}e_u^Te_j)), \\
       & \mathcal{L}_{O} = -\sum_{u = 1}^M\sum_{i \in \mathcal{N}_u}\sum_{j \notin \mathcal{N}_u} ln(\sigma(e_u^Te_i - e_u^Te_j)),
    \end{aligned}
    \label{losses}
\end{equation}
where $\mathcal{N}_u$ is a set of items with which the user $u$ has interacted and $\sigma(x) = \frac{1}{1 + e^{-x}}$ is the sigmoid function.

\vspace{-3pt}
\section{Addressing Warm Start}
\label{method}
As could be seen in Subsection \ref{ultra}, the UltraGCN model \cite{mao2021ultragcn} is similar to the matrix factorization model \ref{svd}, where the user embedding matrix $B_UU$ in UltraGCN is similar to $U$ in SVD, and $B_IV$ is similar to $V\Sigma$. Therefore, to update the user representations, we can utilize a similar optimization problem \eqref{fold}:
\begin{equation}
    \|a_u^T - \beta_ue_u^TV^TB_I\|_2^2 \rightarrow \operatorname*{min}_{e_u}.
    \label{task2}
\end{equation}
Similarly to \eqref{fold}, we compute the derivate of \eqref{task2} and set it equal to zero. The updated warm user embedding equal to:
\begin{equation}
    e_u^T = \frac{1}{\beta_u}(a_u^TB_I^{-1})V^+,
    \label{ufold}
\end{equation}
where $B_I^{-1}$ is the diagonal matrix with $\frac{1}{\beta_i}$ on the diagonal and $V^+$ is the pseudo-inverse matrix $V^+ = A\Sigma_V^{-1}B^T$, where $V = A\Sigma_VB^T$ is the truncated SVD decomposition of the matrix $V$. We employ the pseudo-inverse of the initial matrix V in equation \eqref{fold}, as the orthogonality condition applicable in Singular Value Decomposition does not hold for matrix V \cite{klema1980singular}. 

Due to the fact that the item embeddings $V$ do not change in our Folding-In procedure, the matrices $V^+$ and $B_I^{-1}$ could be precomputed once and then reused for each embedding update. For this reason, the complexity of time and memory is equal to $O(Nd)$, where $c = a_u^TB_I^{-1}$ has $O(N)$ and $e_u^T = \frac{1}{\beta_u}cV^+$ has $O(Nd)$ complexity with only one matrix-vector multiplication. As a result, \emph{our linear approach exhibits linear scalability with respect to catalogue size}. Moreover, due to the utilization of the history of only one user, the method has a small impact on popularity bias \cite{abdollahpouri2019popularity}.

In contrast, conventional stochastic gradient descent-based approaches compute the gradient 
\begin{equation}
    \nabla\mathcal{L}_{e_u} = 2\beta_uV^T(B_I(\beta_uB_I(Ve_u)-a_u))
    \label{sgd}
\end{equation} 
$k$ times. Therefore, the complexity is $O(Ndk)$ with two matrix-vector multiplications for each gradient computation. Therefore, to improve the quality of the SGD-based method recommendations, different heuristics and interactions of other users are used. For instance, to improve the quality of recommendations, the final warm user embedding is computed as $e_u = \mu e_m + (1 - \mu) e_u$, where $e_m = \frac{1}{M}\sum_{i = 1}^MU_i$ is the mean user embedding, and $\mu \in [0,1]$ is a hyperparameter of the method. This may have a negative effect on the scalability of the SGD-based approach due to additional computations and memory consumptions. 

\vspace{-3pt}
\section{Experimental Setup}
\label{exp}
We evaluated our Folding-In approach against several others across four datasets with a wide range of catalogue sizes from $4\cdot10^3$ to $92\cdot10^3$: MovieLens-1M (ML-1M) \cite{movielens}, Amazon's Beauty and Books \cite{amazon}, and the Million Songs Dataset (MSD) \cite{bertin2011million}. The Books and MSD datasets are preprocessed similarly to work \cite{spivsak2023scalable}. The statistics for these datasets are presented in Table \ref{table:stats}. To ensure a robust evaluation and prevent information leakage \cite{leak}, we split the datasets into three subsets (train, warm, and test) according to their timestamps \cite{splits, splits2}. As a result, the datasets are divided into approximately $80\%$, $10\%$, and $10\%$ proportions. The models are trained in the training subset, our methods are applied in the warm subset, and the performance is tested on the test subset.

In our comparisons, we compare the Folding-In approaches with the PureSVD \cite{puresvd} method, as well as the auto-encoder models: EASER \cite{steck2019embarrassingly} and SANSA \cite{spivsak2023scalable}. These models do not utilize user embeddings; therefore, both models naturally solve the warm start problem. The SANSA is the sparse modification of the EASER model which has significantly faster training and inference time and utilizes notably less memory. We also compute the metrics of recommendations for the commonly used graph-based recommender system \cite{he2020lightgcn} which uses the simplification of graph convolution \cite{kipf2016semi}. Additionally, we analyse various techniques for updating warm user representations in the UltraGCN model\cite{mao2021ultragcn}, including Zero (using initial model without Folding-In), Mean -- the mean user embedding for each warm user, SGD-based approaches \cite{sgd}, Full model retraining, and our fast Folding-In method (Linear). Moreover, the Full retraining of the UltraGCN model \cite{mao2021ultragcn} shows significantly faster training and inference with better scalability, compared to LightGCN \cite{he2020lightgcn} and EASER\cite{steck2019embarrassingly}. The performance of the models is evaluated using the Hit Rate and NDCG metrics \cite{silveira2019good}.

\vspace{-5pt}
\begin{table}[h]
\caption{Dataset statistics.}
\vspace{-2pt}
	\label{table:stats}
	\footnotesize
	\centering
        \scalebox{1.02}{
	\begin{tabular}{l|cccc}
		\toprule
  
		\textsf{Statistic} 
         & Users & Items  & Actions &  Density \\
		\midrule
            ML-1M & 6,040 & 3,706 & 1,000,209 & 5.43\% \\ 
            Beauty & 22,363 & 12,101 & 198,502 & 0.073\% \\ 
            Books & 632,458 & 91,599 & 35,918,135 & 0.062\% \\
            MSD & 571,355 & 41,140 & 33,861,510 & 0.144\%\\
		\bottomrule
	\end{tabular}}
    \vspace{-6pt}

\end{table}

\vspace{-3pt}
\section{Results}
\label{res}

In this section, we present the results of our Linear Folding-In approach. The performance of our models is shown in Table \ref{tab:results}. As demonstrated, our Folding-In approach significantly outperforms the SGD-based method in recommendation quality, achieving speeds up to $30\times$ faster while providing superior recommendations. This improvement can be attributed to the use of the exact solution, as opposed to the approximations made by Stochastic Gradient Descent. Furthermore, due to the reduced number of hyperparameters compared to SGD-based approaches, identifying the optimal configuration for the method is considerably faster.

\begin{figure}
    \centering
    \includegraphics[width=0.81\linewidth]{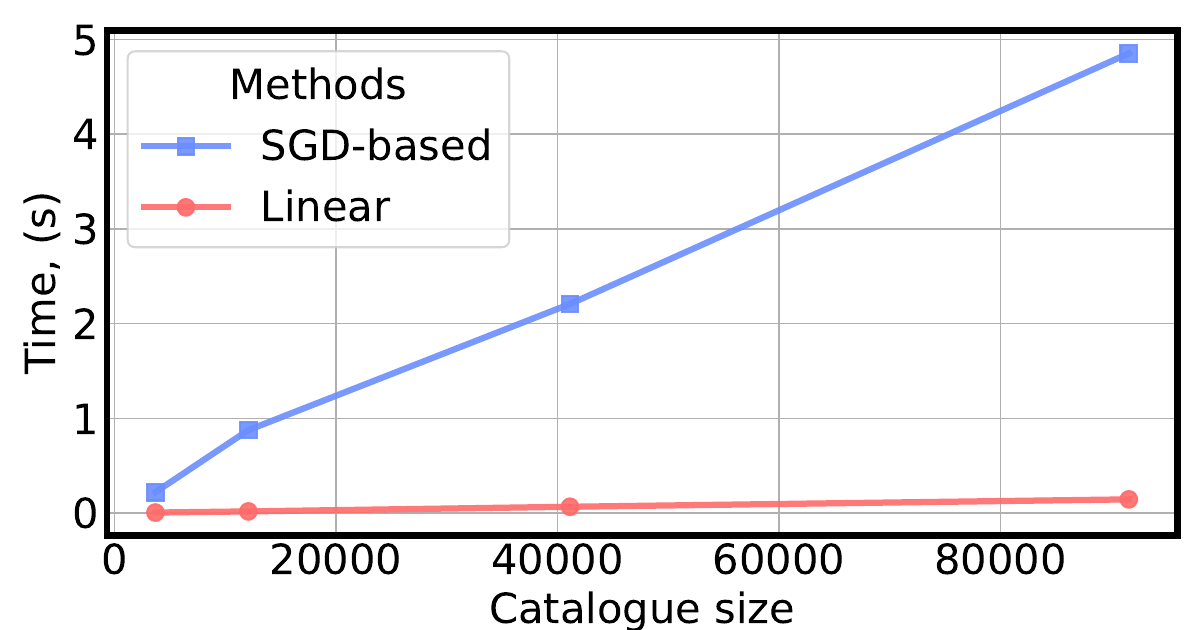}
    \vspace{-8pt}
    \caption{The comparison of time of updating warm user embedding for different catalogue sizes for SGD-based and our Linear approach. The points from the left to the right correspond to ML-1M, Beauty, MSD and Books datasets.
    }
    \label{fig:time}
\end{figure}

\begin{table*}[h]
    \centering
    \caption{Performance comparison of different recommender approaches across datasets. Best metric values are highlighted in \textbf{bold} and second best values are \underline{underlined}. H -- HR metric, N -- NDCG metric, sec/user  -- time in seconds of updating personal recommendations per user. Full retrain approach excluded from comparison due to inapplicable long training time.}
    \vspace{-4pt}
    \scalebox{0.91}{\begin{tabular*}{\textwidth}{@{\extracolsep\fill}c|c|cccc||c|cccc}
        \toprule
        Dataset & Metric & PureSVD & EASER & SANSA & LightGCN & Full & Zero & Mean & SGD & Linear (our) \\
        \midrule
        ML-1M
        & H@5 & 0.0295 & 0.0324 & 0.0336 & 0.0354 & 0.0362 & 0.0341 & 0.0320 & \underline{0.0354} & \textbf{0.0357} \\
        & H@10 & 0.0592 & 0.0605& 0.0612 & 0.0678 & 0.0681 & 0.0625 & 0.0593 & \underline{0.0667} & \textbf{0.0675} \\
        & N@5 & 0.0254 & 0.0266 & 0.0274 & 0.0288 & 0.0290 & 0.0283 & 0.0270 & \underline{0.0285} & \textbf{0.0287} \\
        & N@10 & 0.0315 & 0.0341 & 0.0345 & 0.0370 & 0.0375 & 0.0357 & 0.0338 & \underline{0.0367} & \textbf{0.0369} \\
        \cmidrule{2-11}
        & sec/user & 45.32 & 87.34 & 6.34 &1629 & 729 & -- & -- & \underline{0.223} & \textbf{0.006} \\
        \midrule
        Beauty
        & H@5 & 0.0230 & 0.0237 & 0.0227 &0.0237 & 0.0241 & 0.0228 & 0.0221 & \underline{0.0234} & \textbf{0.0236} \\
        & H@10 & 0.0436 & 0.0448 & 0.0425 & 0.0448 & 0.0453 & 0.0428 & 0.0421 & \underline{0.0441} & \textbf{0.0445} \\
        & N@5 & 0.0158 & 0.0165 & 0.0164 & 0.0171 & 0.0172 & 0.0166 & 0.0159 & \underline{0.0164} & \textbf{0.0165} \\
        & N@10 & 0.0247 & 0.0251 & 0.0243 & 0.0251 & 0.0254 & 0.0245 & 0.0240 & \underline{0.0249} & \textbf{0.0253} \\
        \cmidrule{2-11}
        & sec/user & 113.56 & 205.86 & 13.45 & 2744 & 1175 & -- & -- & \underline{0.872} & \textbf{0.018} \\
        \midrule
        Books
        & H@5 & 0.0265&  0.283 & 0.0291 & 0.0316 & 0.0318 & 0.0301 & 0.0284 & \underline{0.0307} & \textbf{0.0311} \\
        & H@10 & 0.0632&  0.0684 & 0.0698 & 0.0782 & 0.0786 & 0.0748 & 0.0701& \underline{0.0766} & \textbf{0.0774} \\
        & N@5 & 0.0186&  0.0193 & 0.0201 & 0.0216 & 0.0218 & 0.0209 & 0.0191 & \underline{0.0212} & \textbf{0.0215} \\
        & N@10 &  0.0332&  0.352 & 0.361 & 0.0391 & 0.0396 & 0.0379 & 0.0343 & \underline{0.0386} & \textbf{0.0390} \\
        \cmidrule{2-11}
        & sec/user & 132.53 & 232.45 & 64.43 & 32465 & 11813 & -- & -- & \underline{4.852} & \textbf{0.145} \\
        \midrule
        MSD
        & H@5 & 0.0273&  0.305 & 0.0311 & 0.0331 & 0.0337 & 0.0326 & 0.0318 & \underline{0.0329} & \textbf{0.0334} \\
        & H@10 & 0.0672&  0.0703 & 0.0714 & 0.0775 & 0.0783 & 0.0759 & 0.0751& \underline{0.0765} & \textbf{0.0772} \\
        & N@5 & 0.0201&  0.0214 & 0.0223 & 0.0242 & 0.0245 & 0.0231 & 0.0221 & \underline{0.0235} & \textbf{0.0240} \\
        & N@10 &  0.0393&  0.405 & 0.413 & 0.0427 & 0.0432 & 0.0413 & 0.0401 & \underline{0.0421} & \textbf{0.0426} \\
        \cmidrule{2-11}
        & sec/user & 101.87 & 176.01 & 54.43 & 26746 & 9745 & -- & -- & \underline{2.207} & \textbf{0.067} \\
        \bottomrule
    \end{tabular*}}
    \label{tab:results}
\end{table*}
To illustrate the linear scalability of our method, we examine the relationship between the updating time per user and the catalogue size. In this experiment, we set the embedding size $d = 32$ for both the Linear and SGD-based approaches to empirically assess the scalability. As shown in Figure \ref{fig:time}, our Linear approach exhibits linear scalability with a $O(N)$ complexity.

\begin{figure}
    \centering
    \includegraphics[width=0.81\linewidth]{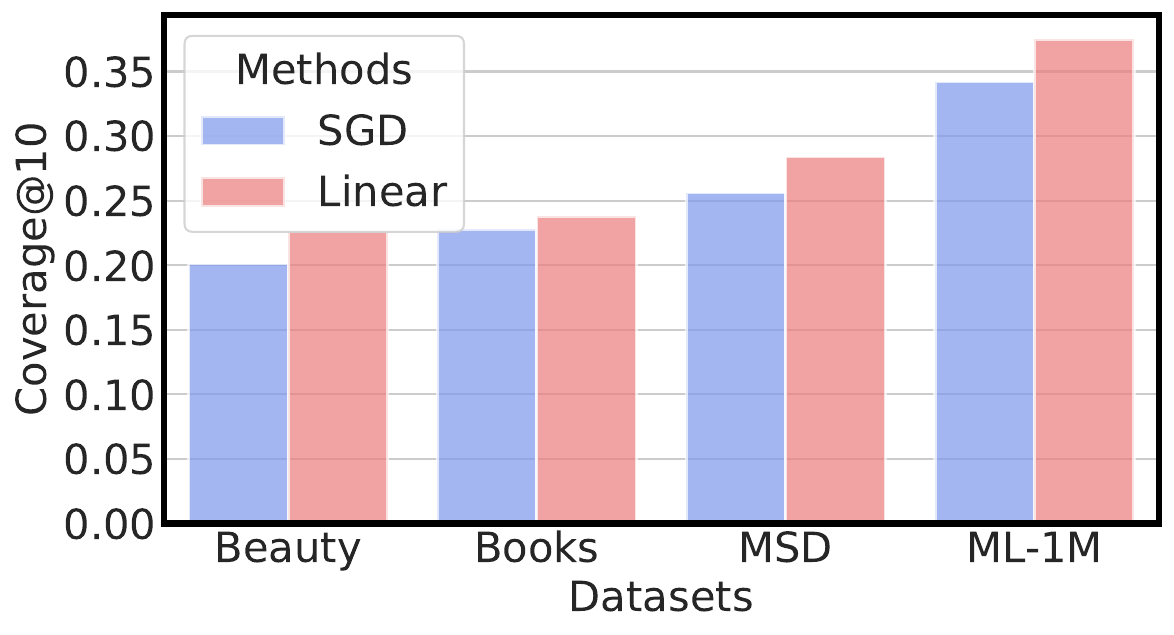}
    \vspace{-8pt}
    \caption{The comparison of coverage@$10$ on four datasets for the SGD-based and our Linear Folding-In approaches.}
    \label{fig:cov}
\end{figure}

Moreover, by utilizing only the history of warm users, our Linear Folding-In approach effectively mitigates popularity bias \cite{abdollahpouri2019popularity}. In contrast, the tuned conventional SGD-based approach may use heuristics and the histories of other users (Section \ref{method}), which can adversely affect popularity bias in personalized recommendations. To investigate this effect, we compare the coverage metrics of the SGD-based and our Linear approaches. As shown in Figure \ref{fig:cov}, the coverage of SGD-based recommendations is notably lower than that of our Linear approach across all datasets. Therefore, the diversity of recommendations of our Linear method is also greater \cite{abdollahpouri2019popularity}.

\vspace{-3pt}
\section{Conclusion}
\label{conc}
In this work, we developed a time- and resource-efficient scalable Folding-In approach to effectively address the warm start problem in UltraGCN-like recommender systems. Our Linear method achieves up to a $30\times$ speedup over conventional SGD-based approaches while also enhancing the quality of personalized recommendations and demonstrating linear scalability with respect to catalogue size. Furthermore, our approach is less susceptible to popularity bias compared to traditional methods, resulting in more diverse recommendations. 

\vspace{-4pt}
\begin{acks}
Support from the Basic Research Program of HSE University is gratefully acknowledged. The calculations
were performed in part through the computational resources of HPC facilities at HSE University \cite{kostenetskiy2021hpc}.
\end{acks}

\balance
\printbibliography
\section*{GenAI Usage Disclosure}
In preparing this paper, we used DeepSeek and ChatGPT to identify and correct spelling errors, typos, and improve the clarity of the text. No AI tools were used for data analysis, experimentation, formulation of conclusions, and mathematical formulations. The research methodology, results, and all other technical aspects of the work were developed by the authors of the article without using GenAI. We utilize DeepSeek and ChatGPT only for proofreading and stylistic issues, while we maintain the full academic integrity and originality of our research.

\end{document}